\documentstyle[aps,prl,multicol,amsfonts]{revtex} 
\renewcommand{\narrowtext}{\begin{multicols}{2} \global\columnwidth20.5pc}
\renewcommand{\widetext}{\end{multicols} \global\columnwidth42.5pc}
\begin{document}

\bibliographystyle{prsty}

\draft

\title{Nonperturbative Renormalization Group Function for Quantum Hall
  Plateau Transitions Imposed by Global Symmetries}

\author{Nobuhiko Taniguchi}

\address{
  Department of Physical Electronics, %
  Hiroshima University, Higashi-Hiroshima 739-8527, Japan}

\date{October 21, 1998}

\maketitle

\begin{abstract}
  As a unified theory of integer and fractional quantum Hall plateau
  transitions, a nonperturbative theory of the two-parameter scaling
  renormalization group function is presented.  By imposing global
  symmetries known as ``the law of corresponding states'', we seek a
  possible form of renormalization group flows.  Asking for consistency
  with the result from weak-localization perturbation theory, such
  restriction is so intense that we can analytically determine its
  concrete form.  Accordingly, the critical exponent $\nu$ and the
  irrelevant scaling index $y$ are obtained analytically and turn out to
  be irrational. Their values ($\nu\approx 2.1$, $y\approx 0.3$) agree
  favorably well with experiments and numerics.
\end{abstract}%


\pacs{Suggested PACS: 73.40.Hm, 68.35.Rh, 64.60.Ak}

\narrowtext

Extensive studies of years have revealed that electronic systems in strong
magnetic field not only possess precisely quantized Hall conductance, but
also exhibit rich structures of stable phases (quantum Hall states) and
quantum phase transitions.  While understanding on various bulk and edge
properties of stable phases in integer and fractional regimes have made a
considerable progress~\cite{PrangeSarma}, our understanding is still
limited on the nature of quantum Hall transitions, namely, quantum phase
transitions occurring between Hall plateaus.
In particular, one of the theoretical challenges is to obtain analytically
the value of the critical exponent which is observed ubiquitously for
integer and fractional quantum Hall transitions both in experiments and
numerics~\cite{WeiKochHuckestein}.

In this Letter, we provide a new and fully nonperturbative treatment for
quantum Hall transitions, based on the two-parameter scaling
theory~\cite{Khmelnitskii83,Pruisken}.  In search of a unified scaling
description of quantum Hall transitions both in integer and fractional
regimes, we require renormalization group (RG) flows to manifest discrete
symmetries suggested by ``the law of
corresponding states''~\cite{Kivelson92}.
Satisfying such symmetric constraint, an explicit form of RG function is
carefully constructed to be consistent with the results from
weak-localization theory.  The obtained result enables us to find
analytically the critical exponent $\nu$ of the correlation length and the
irrelevant scaling index $y$ by expanding around the critical point.

The present work shares a lot in the spirit with recent remarkable
progress in $N=2$ supersymmetric (SUSY) Yang-Mills theory, initiated by
Seiberg and Witten~\cite{SeibergWitten}.  Subsequently, the
nonperturbative form of RG function was determined
exactly by utilizing discrete symmetries present in the
model~\cite{Minahan96,Ritz98}. We should keep in mind, however, a crucial
difference between these two models: the theory of quantum Hall
transitions {\em does not} possess $N=2$ SUSY, as is seen from the
large-conductance expansion of the standard weak-localization theory, or a
corresponding 2D nonlinear $\sigma$-model~\cite{Wegner89},
\begin{equation}
   {d\ln\sigma_{xx}\over d\ln L} = - {1\over 2\pi^2 \sigma_{xx}^2} - {C\over
     \sigma_{xx}^4} +   \cdots,
\label{eq:NLsigma}
\end{equation}
where $C$ is some constant.  It shows, in contrast with the Yang-Mills
theory, that higher-loop corrections exist in the theory, hence
holomorphic property of $N=2$ SUSY cannot be used to investigate the
problem.  Notwithstanding it is emphasized that what is important below is
the presence of infinite-dimensional discrete
symmetries~\cite{CardyShapere}, not $N=2$ SUSY.

Discrete symmetries that we are going to utilize are the ones proposed in
Ref.~\cite{Kivelson92}, to account for the observed superuniversality of
integer and fractional quantum Hall transitions~\cite{WeiKochHuckestein},
as well as the geometry of the global phase diagram~\cite{Kravchenko95}.
Symmetrical transformations are composed of equivalence by Landau level
addition and/or by flux-attachment (as well as particle-hole
transformation which we will not use explicitly below).  
By introducing the complexified conductance $\tau \equiv \sigma_{xy} +
i\sigma_{xx}$ (measured in the unit of $e^2/h$), the content of this
equivalence on the parameter space $(\sigma_{xy},\sigma_{xx})$ can be
neatly related to modular transformations: translation $\tau \to T\tau =
\tau + 1$, and inversion $\tau \to S\tau = -1/\tau$~\cite{Lutken93}.  In
fact, elementary transformations of the laws of corresponding states are
given by
\begin{eqnarray*}
  T&&:\: \tau \to  \tau+1 \quad \mbox{(Landau-level-addition)},\\
  ST^2 S&&:\: \case{1}{\tau} \to  \case{1}{\tau} -
  2\quad\mbox{(flux-attachment)}.
\end{eqnarray*}
Hence the relevant symmetry is a subgroup of $SL(2,{\Bbb Z})$ generated by
$T$ and $ST^2S$.  This group is customarily designated as $\Gamma_{0}(2)$
in literatures~\cite{MathBooks}, and we will follow this convention below.
The critical point of the integer quantum Hall transitions of the lowest
Landau level is found to be $(\sigma_{xy}^{c},\sigma_{xx}^{c}) =
(\frac{1}{2},\frac{1}{2})$ or $\tau_{c} = \frac{1+i}{2}$~\cite{Huo93}, and
so are expected all the equivalent points $\gamma\tau_{c}$ by
transformations $\gamma\in \Gamma_{0}(2)$.  Of course, such equivalence
exists only approximately in any real system.  We, however, take this as
exact symmetry, and will explore its consequence on RG flows and critical
exponents, without asking its microscopic origin.
It is remarked that both the duality and the semicircle
laws~\cite{ShaharRuzin} were already presented implicitly in
Ref.~\cite{Lutken93}, and RG flows of $\Gamma_{0}(2)$ will give a natural
explanation for it.

As a phenomenological theory for {\em integer} quantum Hall transitions,
the two-parameter scaling theory~\cite{Khmelnitskii83,Pruisken} has been
quite successful to analyze data from experiments and
numerics~\cite{WeiKochHuckestein}, though it fails to predict the values
of the critical exponents theoretically so far.  Plateau transitions are
believed to be the second-order, so that $\beta$-function is expected to
have a linear deviation $\delta\tau$ around the transition points, say,
$\tau_{c}$.  By introducing the complexified RG function
\begin{equation}
  \beta(\tau,\overline{\tau}) = \frac{d\tau}{d\ln L} =
  \frac{d\sigma_{xy}}{d\ln L} + i \frac{d\sigma_{xx}}{d\ln L},
\end{equation}
the critical behavior is encoded in the form, 
\begin{eqnarray}
 \beta(\tau,\overline{\tau})  &&\approx \nu^{-1} {\rm Re}\,\delta\tau -
  y \;{\rm Im}\,\delta\tau + \cdots .
\label{eq:scaling}
\end{eqnarray}
Here we have attached $\tau$ and $\overline{\tau}$ as arguments of $\beta$
to remind that it is non-analytical, as already suggested in
Eq.~(\ref{eq:NLsigma}).  The critical exponent $\nu$ characterizes the
divergence of the correlation length, and $y$ corresponds to the largest
irrelevant scaling index.  Experimental and numerical observed values
consistently show $\nu\approx 2.3 \pm 0.1$ and $y\approx
0.4$~\cite{WeiKochHuckestein,comment-eta}.
Theoretical attempts to derive such $\beta$-function were made in
Ref.~\cite{Pruisken} by using instanton-approximation of nonlinear
$\sigma$-model with a topological term.  In the complex notations, the
result was
\begin{equation}
  \beta(\tau,\overline{\tau}) \approx -{i/2\pi^2 \over {\rm Im}\,\tau} - {i\, C
    \over ({\rm Im}\,\tau)^3}- i D\,({\rm Im}\,\tau)^{n}\, e^{-2i\pi
    \overline{\tau}}, 
  \label{eq:large-tau}
\end{equation} 
where $D$ is a constant, and we add as the second term the four-loop
contribution from Eq.~(\ref{eq:large-tau}).  Although actual values of
$\nu$ and $y$ cannot be evaluated from here, Eq.~(\ref{eq:large-tau})
tells us the (nonperturbative) asymptotics of RG function for
$\tau\to+i\infty$~\cite{comment-Pruisken}.

Discrete symmetries $\Gamma_{0}(2)$ imposed by the laws of
corresponding states, constrain possible RG flows considerably.  It
means, for instance, knowing the form Eq.~(\ref{eq:scaling}) around
$\tau_{c}$ immediately enables us to have RG flows around {\it all} the
equivalent transition points $\gamma\tau_{c}$ for $\gamma\in
\Gamma_{0}(2)$.  Mathematically, it requires that for the symmetry
transformation $\tau\to \gamma\tau = (a\tau+b)/(c\tau+d)$,
$\beta$-function should satisfy the
functional relation~\cite{Burgess97,Damgaard97},
\begin{eqnarray}
  \beta(\gamma \tau, \gamma\overline{\tau}) && = {d(\gamma\tau) \over
    d\ln L}= (c\tau+d)^{-2}\beta(\tau,\overline{\tau}).  
\label{eq:beta-rel}
\end{eqnarray}
Functions with such property are termed as automorphic functions with
weight $(-2,0)$~\cite{MathBooks}.  In general, automorphic functions with
weight $(k,k')$ are defined similarly by the functional relation
\begin{equation}
  F(\gamma \tau,\gamma\overline{\tau}) =
  (c\tau+d)^{k}(c\overline{\tau}+d)^{k'} F(\tau,\overline{\tau}).  
\label{eq:automorphism}
\end{equation}

Basic building blocks to construct automorphic functions are the
analytical function $f(\tau)$ which remains invariant under
$\Gamma_{0}(2)$, and the Eisenstein series $G_{2}^{(ij)}$, which turn out
to be non-analytical~\cite{MathBooks}.  They are defined by
\begin{eqnarray}
 &&f(\tau) = -{{\vartheta_{3}^{4}(0|\tau) \vartheta_{4}^{4}(0|\tau) \over
\vartheta_{2}^{8} (0|\tau)} },
\label{def:f}\\
  &&G_{2}^{(ij)}(\tau,\overline{\tau}) = \lim_{s\to
    0}\mathop{\sum\nolimits'}_{(m,n)\equiv 
    (ij)}{1\over (m\tau +n)^2 |m\tau+n|^{s}},
\end{eqnarray}
where $\vartheta_{a}(u|\tau)$ are elliptic theta
functions~\cite{Abramowitz74}, and the summation on the second line is
taken over two integers $(m,n)$ satisfying $m\equiv i$ and $n\equiv j
\bmod{2}$ except for $(0,0)$.  Among them, $G_{2}^{(00)}$ and
$G_{2}^{(01)}$ become automorphic with weight $(2,0)$ under
$\Gamma_{0}(2)$.
For our purpose below, it is instead more convenient to use the following
pair of linear combinations~\cite{comment-Burgess},
\begin{mathletters}
\begin{eqnarray}
  \phi_{0}(\tau,\overline{\tau}) &&\equiv {2\over\pi}
  \left[G_{2}^{(01)}(\tau,\overline{\tau}) - 3
    G_{2}^{(00)}(\tau,\overline{\tau})\right] \nonumber \\ 
  &&={1\over {\rm Im}\,\tau}+ 16\pi \sum_{n=1}^{\infty} {n\,
    e^{2in\pi\tau}\over  {1- e^{4in\pi\tau}}},\\
  \phi_{1}(\tau) &&\equiv {6 \over
    \pi^2}\left[G_{2}^{(01)}(\tau,\overline{\tau}) -
    G_{2}^{(00)}(\tau,\overline{\tau}) \right] \nonumber\\ 
  &&=1+24 \sum^{\infty}_{n=1}{n\, e^{2in\pi\tau}\over
    1+e^{2in\pi\tau}} .
\end{eqnarray}
\end{mathletters}%
Our choice of $\{\phi_{0},\phi_{1}\}$ is made to sort out different
analyticity and different $\tau\to +i\infty$ asymptotic,
\begin{mathletters}
\begin{eqnarray}
  &&\phi_{0}(\tau,\overline{\tau}) \approx {1\over {\rm Im}\,\tau}+ 16\pi
  e^{2i\pi\tau} + 32\pi e^{4i\pi\tau} +  \cdots, \\
  &&\phi_{1}(\tau) \approx 1+24 e^{2i\pi\tau} + 24 e^{4i\pi\tau} + \cdots .
\end{eqnarray}
\end{mathletters}%
$\phi_{1}$ is holomorphic and related to $f$ by $\phi_{1}=(i/2\pi) f'/f$.
It is easy to see that the inverse and the complex conjugate of
$\phi_{0,1}$ behave automorphically with weight $(-2,0)$ and $(0,2)$,
respectively.

We use $f$ and $\phi_{0,1}$ to construct $\beta$-function of quantum Hall
transitions with desired properties,
Eqs.~(\ref{eq:scaling},\ref{eq:large-tau}).  To illustrate the steps, it
is instructive to mention a recent neat derivation~\cite{Ritz98} of RG
function $\beta_{\rm SYM}$ in SUSY Yang-Mills theory, where $N=2$ SUSY
makes a procedure much simpler.
Note that low-energy effective action of such theory enjoys exactly the
same discrete symmetries, $\Gamma_{0}(2)$, though the phase structure
differs.  By knowing its holomorphy and the asymptotics $\beta_{\rm
  SYM}(+i\infty) \to 2/i\pi$, it is required
mathematically~\cite{MathBooks} that $(-2,0)$-automorphic function must be
expressed in the following form,
\begin{equation}
  \beta_{\rm SYM}(\tau) = \frac{2}{i\pi}\cdot \frac{1}{\phi_{1}}\cdot {{\cal
      P}(f) \over {\cal Q}(f)},
\label{eq:betaSYM}
\end{equation}
where ${\cal P}$ and ${\cal Q}$ are polynomials with the same coefficients
for the highest order term to ensure ${\cal P}/{\cal Q}\to 1$ for $\tau\to
+i\infty$.  They determine locations of zeros and poles of $\beta_{\rm
  SYM}$, so govern the phase structure.  Putting the other way round, the
forms of ${\cal P}$ and ${\cal Q}$ can be deduced when we have some grasp
of the phase diagram, as is the case of quantum Hall systems.  For
instance, the correct answer ${\cal P} = f-\frac{1}{4}$ and ${\cal Q}=f$
could be obtained easily once we knew $\beta_{\rm SYM}(\tau_{c})=0$ and
$\beta_{\rm SYM}(0)=\infty$.

We use a similar reasoning to construct the $\beta$ function for the
quantum Hall transitions.  An key observation is that non-holomorphic
function $\beta(\tau,\overline{\tau})$ which behaves asymptotically as in
Eq.~(\ref{eq:large-tau}) simultaneously satisfies the automorphic
condition of weight $(-2,0)$.  Possibility for automorphic functions to
have power-law $({\rm Im}\,\tau)^n$ is very much limited.  First is to notice
that ${\rm Im}\,\tau$ itself is automorphic with weight $(-1,-1)$.  Another
possibility is, similarly to $G_{2}^{(ij)}$, that analytical continuation
of Eisenstein series into $k+k\le 2$ may produce terms
$({\rm Im}\,\tau)^{1-k-k'}$~\cite{MathBooks}.  It is also seen that the
instanton contribution of Eq.~(\ref{eq:large-tau}) appears as a function
of $\overline{\tau}$.
With these observations in mind, to satisfy
Eqs.~(\ref{eq:scaling}--\ref{eq:beta-rel}), we choose to write down
$\beta$-function as
\begin{equation}
  \beta(\tau,\overline{\tau}) ={-i\over 2\pi^2}\cdot 
  {\overline{\phi}_{0}\over \left(\phi_{1}+ A \phi_{0}\right)
  \left(\overline{\phi}_{1}-A\overline{\phi}_{0}\right)}\cdot {{\cal
  P}(f,\overline{f})\over {\cal Q}(f,\overline{f})},
\label{def:beta}
\end{equation}
where ${\cal P}$ and ${\cal Q}$ are polynomials of $f$ and
$\overline{f}=f(-\overline{\tau})$, satisfying ${\cal P}/{\cal Q}\to 1$
for $\tau \to +i\infty$.  It is remarked that the possibility $\beta
\propto \overline{\phi}_{0}$ was suggested earlier in
Ref.~\cite{Burgess97}.  Though we cannot prove that Eq.~(\ref{def:beta})
is the only possible choice compatible with
Eqs.~(\ref{eq:scaling}--\ref{eq:beta-rel}), requirement is so strict that
possibility seems very much restricted.

Based on Eq.~(\ref{def:beta}), we now turn our attention to the critical
behavior around $\tau_{c}$.  From Eq.~(\ref{eq:scaling}), a linear
deviation is required.  On the other hand, it is readily seen that both
$\phi_{0}$ and $\phi_{1}$ vanish at $\tau_{c}$~\cite{Taniguchi-preprint}.
So to ensure the form of Eq.~(\ref{eq:scaling}), it is necessary to remove
such a pole by choosing an appropriate form of polynomial ${\cal
  P}(f,\overline{f})$.
To do so, we first expand $f$ and $\phi_{0,1}$ around $\tau_{c}$:
\begin{mathletters}
\begin{eqnarray}
  &&\phi_{0}(\tau,\overline{\tau}) \approx -2i c_{0} \delta\tau -2i
    \overline{\delta\tau}+\cdots, \\
  &&\phi_{1}(\tau) \approx -{4i\over\pi}c_{0} \delta\tau + \cdots\\
  && f(\tau) \approx {1\over 4} -c_{0}(\delta\tau)^2 +\cdots, 
\end{eqnarray}
\end{mathletters}%
Evaluation of $c_{0}$ may be straightforward but rather
laborious~\cite{Taniguchi-preprint}.  We did it by reducing into elliptic
functions of the lemniscate case~\cite{Abramowitz74}, to produce
\begin{equation}
 c_0 =  {\Gamma^8({1\over 4}) \over 64 \pi^4} \approx  4.7892 .
\label{def:c0}
\end{equation}
Note that there is no linear deviation in $f$, so as to make it possible
to remove the pole at $\tau_{c}$ by ${\cal P}$, the denominator
\begin{eqnarray*}
  &&(\phi_{1}+A\phi_{0})(\overline{\phi}_{1}-A\overline{\phi}_{0}) \\
  &&\quad \approx -4 c_{0} A
  (A+{2\over\pi}) (\delta\tau)^2 - 4 c_{0} A (A-{2\over
      \pi})(\overline{\delta\tau})^2 \\
    &&\qquad - 4\left[A^2(1+c_{0}^2)-{4 c_{0}^2 \over
      \pi^2}\right]  
    (\overline{\delta\tau})(\delta\tau) 
\end{eqnarray*}
is required to have no cross term $(\overline{\delta\tau})(\delta\tau)$.
This determines the value of $A$ as
\begin{equation}
    A = {2 c_{0} \over \pi \sqrt{1+c_{0}^2}}.  
    \label{def:A}
\end{equation}
(The sign of $A$ determines the direction of RG flows as will be seen
below, so $A$ must be positive.)  To cancel out the remaining terms,
${\cal P}$ need to include the factor proportional to
\begin{equation}
4 A \left[
  (A+{2\over\pi}) f + (A-{2\over\pi})\overline{f}-{A\over 2}\right].
\end{equation}
In addition, it is expected on physical grounds that there is no other
zeros ({\it i.e.}, transition points) in $\beta$-function other than
$\tau_{c}$ and equivalent points.  Hence ${\cal P}$ must be a linear
function of $f$ and $\overline{f}$, and we choose as
\begin{equation}
  {\cal P}(f,\overline{f}) = f + \overline{f} + {2(f-\overline{f})\over
      \pi A} - {1\over 2}.
    \label{def:P}
\end{equation}
To determine ${\cal Q}$, we recall ${\cal P}/{\cal Q}\to 1$ for $\tau\to
+i\infty$, so ${\cal Q}(f,\overline{f})$ must also be a linear polynomial
of $f$ and $\overline{f}$ with the same coefficients of ${\cal P}$.  In
addition, $\beta$ will diverge at zeros of ${\cal Q}$.  Since only places
where $\beta$ is allowed to diverge are at the points corresponding to the
stable phases (quantum Hall and insulator states), so we conclude that
\begin{equation}
  {\cal Q}(f,\overline{f}) = f + \overline{f} + {2(f-\overline{f})\over
      \pi A} .  
    \label{def:Q}
\end{equation}
By combining Eqs.~(\ref{def:A},\ref{def:P},\ref{def:Q}) with
Eq.~(\ref{def:beta}) we reach our main result of the paper: the explicit
form of $\beta$-function.

The result obtained incorporates all nonperturbative effect, so it has its
full validity even near the critical points.  As a result, we can simply
expand $\beta$-function around $\tau_{c}$ to evaluate the critical
exponents.
\begin{eqnarray}
  \beta(\tau,\overline{\tau}) \approx {c_{0}+1\over
    2A\pi^2}\,{\rm Re}\,\delta\tau - {c_{0}-1\over 2 A \pi^2}\,{\rm
    Im}\,\delta\tau   +\cdots .
\end{eqnarray}
By comparing this with Eq.~(\ref{eq:scaling}), we can readily
read off the values of the critical exponents $\nu$ and $y$,
\begin{mathletters}
\label{eq:result}
\begin{eqnarray}
  &&\nu = {4\pi c_{0} \over (1+c_{0})\sqrt{1+c_{0}^{2}}}\approx
  2.12, \\
  &&y = {(c_{0}-1) \sqrt{1+c_{0}^{2}} \over 4\pi c_{0}}\approx 0.31.
\end{eqnarray}
\end{mathletters}%
where $c_{0}$ is defined by Eq.~(\ref{def:c0}).  

In conclusion, to investigate the transition between quantum Hall
plateaus, we have constructed RG $\beta$-function by requiring discrete
symmetries of the global phase diagram.  A possible form of RG function
was presented explicitly, and the critical exponent $\nu$ and the
irrelevant scaling index $y$ were extracted analytically.  These values
have turned out to be irrational and suggests $\nu\approx 2.12$ and
$y\approx 0.31$.  Although there is still a small discrepancy with
observed values in experiments and numerics ($\nu\approx 2.4$ and
$y\approx 0.4$), they agree fairly well.  It is striking that simple
notion of two-parameter scaling can provide a {\it quantitative}
prediction of the critical behaviors, with the help of such symmetry
consideration.  The implication of irrational critical exponents is not
clear at present, though.  Finally it is also remarked that the present
construction will be of interest in the study Yang-Mills theory as well,
since it provides means to construct RG functions for systems without
$N=2$ SUSY, but with an approximate modular invariant phase diagram.

The author acknowledges the hospitality at ICTP, Trieste, where part of
the work was done during participating in the workshop {\it Disorder,
  Chaos and Interaction in Mesoscopic Systems}.

%



\begin{references}
\bibitem{comment-Pruisken} In his later publication of
  Ref.~\cite{Pruisken}, the result of $n=2$ was obtained.  However, we
  have found it very hard to make such a term compatible with
  Eqs.~(\ref{eq:scaling},\ref{eq:beta-rel}), so it is suspected that this
  power may be an artifact of the replica limit.  Our result
  Eq.~(\ref{def:beta}) suggests the asymptotics with $n=0$.
\end{references}

\widetext

\end{document}